\documentclass[twocolumn,showpacs,aps,prl,superscriptaddress,letterpaper]{revtex4}

\usepackage{graphicx}
\usepackage{dcolumn}
\usepackage{amsmath}
\usepackage{epsfig}
\usepackage{multirow}
\RequirePackage{xspace}
\usepackage{relsize}

\def\beq{\begin{equation}}
\def\eeq{\end{equation}}
\def\bea{\begin{eqnarray}}
\def\eea{\end{eqnarray}}
\def\nn{\nonumber}
\def\sss{\scriptscriptstyle}
\def\roughly#1{\mathrel{\raise.3ex\hbox
{$#1$\kern-.75em\lower1ex\hbox{$\sim$}}}}

\def\bd{B_d^0}
\def\bs{B_s^0}
\def\bdbar{B_d^0}
\def\bsbar{B_s^0}
\def\btod{{\bar b} \to {\bar d}}
\def\btos{{\bar b} \to {\bar s}}

\def\bra#1{\left\langle #1\right|}
\def\ket#1{\left| #1\right\rangle}
\def\fT{f_{\sss T}}
\def\fL{f_{\sss L}}
\def\fTfL{f_{\sss T}/f_{\sss L}}

\begin{document}

\begin{flushleft}
UMISS-HEP-2008-02 \\
UdeM-GPP-TH-08-167 \\
UAB-FT-646\\[10mm]
\end{flushleft}

\title{
\large \bfseries \boldmath
 Final-state Polarization in $\bs$ Decays}

%
\author{Alakabha Datta}\thanks{datta@phy.olemiss.edu}
\affiliation{Dept.\ of Physics and Astronomy, 108 Lewis Hall,
University of Mississippi, Oxford, MS 38677-1848, USA}
\author{David London}\thanks{london@lps.umontreal.ca}
\affiliation{Physique des Particules, Universit\'e de
Montr\'eal,C.P. 6128, succ. centre-ville, Montr\'eal, QC,
Canada H3C 3J7}
\author{Joaquim Matias}\thanks{matias@ifae.es}
\affiliation{IFAE, Universitat Aut\`onoma de Barcelona, 08193
Bellaterra, Barcelona, Spain}
\author{Makiko Nagashima}\thanks{makiko@lps.umontreal.ca}
\author{Alejandro Szynkman}\thanks{szynkman@lps.umontreal.ca}
\affiliation{Physique des Particules, Universit\'e de
Montr\'eal,C.P. 6128, succ. centre-ville, Montr\'eal, QC,
Canada H3C 3J7}

\date{\today}

\begin{abstract}
Certain $\bs\to V_1V_2$ decays ($V_i$ is a vector meson) can
be related by flavor SU(3) symmetry to corresponding $\bd\to
V_3V_4$ decays. In this paper, we show that the final-state
polarization can be predicted in the $\bs$ decay, assuming
polarization measurements of the $\bd$ decay.  This can be
done within the scenario of penguin annihilation (PA), which
has been suggested as an explanation of the unexpectedly
large transverse polarization in $B\to\phi K^*$. PA is used
to estimate the breaking of flavor SU(3) symmetry in pairs of
decays.  Two of these for which PA makes a reasonably precise
prediction of the size of SU(3) breaking are ($\bs \to \phi
\phi, \bd \to \phi K^{0*}$) and ($\bs \to \phi {\bar K}^{0*},
\bd \to {\bar K}^{0*} K^{0*}$). The polarization measurement
in the $\bd$ decay can be used to predict the transverse
polarization in the $\bs$ decay, and will allow a testing of
PA.
\end{abstract}


\maketitle


\section{Introduction}

We consider $B\to V_1V_2$ decays ($V_i$ is a vector
meson). Since the final-state particles are vector mesons,
when the spin of these particles is taken into account, this
decay is in fact three separate decays, one for each
polarization (one longitudinal, two transverse). Naively,
within the standard model (SM), the transverse amplitudes are
suppressed by a factor of size $m_{\sss V}/m_{\sss B}$ ($V$
is one of the vector mesons) with respect to the longitudinal
amplitude. Then one expects the fraction of transverse
decays, $\fT$, to be much less than the fraction of
longitudinal decays, $\fL$.

However, it was observed that these two fractions are roughly
equal in the decay $B\to\phi K^*$: $\fTfL \simeq 1$
\cite{phiK*,phiK0*recent,phiK+*recent} A similar effect was
later seen in $B \to \rho K^*$ decays \cite{rhoK*}.

If one goes beyond the naive SM, there are two explanations
\cite{DGLNS} which account for this ``polarization puzzle.''
The first is penguin annihilation (PA) \cite{Kagan}.
$B\to\phi K^*$ receives penguin contributions, ${\bar b}
{\cal O} s {\bar q} {\cal O} q$, where $q=u,d$ (${\cal O}$
are Lorentz structures, and color indices are suppressed).
With a Fierz transformation, these operators can be written
as ${\bar b} {\cal O}' q {\bar q} {\cal O}' s $. A gluon can
now be emitted from one of the quarks in the operators, and
can then produce a pair of $s, \bar{s}$ quarks.  These then
combine with the ${\bar s}, q$ quarks to form the final
states $\phi K^{*+}~(q=u)$ or $\phi K^{*0}~(q=d)$. These are
annihilation contributions. Normally such terms are expected
to be small as they are higher order in the $1/m_b$
expansion, and thus ignored.  However, within QCD
factorization (QCDf) \cite{BBNS}, it is plausible that the
coefficients of these terms are large \cite{Kagan}. (Within
perturbative QCD \cite{pQCD}, the penguin annihilation is
calculable and can be large, though it is not large enough to
explain the polarization data in $B\to\phi K^*$ \cite{LM}.)

In QCDf, due to the appearance of endpoint divergences, PA
is not calculable, but is modeled \cite{BBNS}. These
divergences are regulated with a cut-off, introducing several
arbitrary parameters. There is therefore an enormous
uncertainty in the size of the PA amplitude as one varies
these unknown parameters within certain chosen limits
\cite{Benekeetal}. 

It is also possible within QCDf that the transverse
amplitudes receive significant contributions from
perturbative rescattering from charm intermediate
states. However, the transverse amplitudes could be purely
dominated by PA. In this paper we explore the consequences of
the scenario in which PA contributions are large and dominant
to see what type of testable predictions result.

The second SM explanation is rescattering \cite{soni, SCET}.
The idea is that nonperturbative rescattering effects
involving charm intermediate states, generated by the
operator ${\bar b} {\cal O}' c {\bar c} {\cal O}' s$, can
produce large transverse polarization in $B\to\phi K^*$.  A
particular realization of this scenario is the following
\cite{soni}.  Consider the decay $B^+ \to D_s^{*+} {\bar
D}^{*0}$. Since the final-state vector mesons are heavy, the
transverse polarization can be large.  The state $D_s^{*+}
{\bar D}^{*0}$ can now rescatter to $\phi K^{*+}$. If the
transverse polarization $T$ is not reduced in the scattering
process, this will lead to $B^+\to\phi K^{*+}$ with large
$\fTfL$. (A similar rescattering effect can take place for
$\bdbar\to\phi K^{*0}$.)

It is important to test these explanations in order to
determine whether new physics is or is not present. The
polarization puzzle has been mainly seen in $\btos$
transitions.  However, if PA or rescattering is the true
explanation, one also expects to observe large $\fTfL$ in
$\btod$ decays. In Ref.~\cite{DGLNS} such decays were
discussed, and it was observed that the most promising
transitions were those which are dominated by penguin
amplitudes. The $\btos$ and $\btod$ penguin decays are
\bea
\btos s {\bar s} ~{\rm and}~ \btos d {\bar d} & : &
\bd \to \phi K^{0*} \nn\\
& & \bs \to \phi \phi, K^{0*} {\bar K}^{0*} \nn\\
& & B^{\sss +} \to \phi K^{{\sss +}*}, \rho^{\sss +} K^{*0} \nn\\
\btod s {\bar s} ~{\rm and}~ \btod d {\bar d} & : &
\bd \to {\bar K}^{0*} K^{0*} \nn\\
& & \bs \to \phi {\bar K}^{0*} \nn\\
& & B^{\sss +} \to K^{{\sss +}*} {\bar K}^{0*}
\label{list}
\eea
(Decays which also receive tree contributions are not
included in the current analysis.)

Now, all of these decays are the same under flavor SU(3),
which treats $d$, $s$ and $u$ quarks as equal. The idea is
that, given a measurement of the polarization in one decay,
one can predict the polarization in another decay using PA or
rescattering. However, in relating the two decays, the effect
of SU(3) breaking must be included. We can relate the
transverse amplitudes of SU(3)-related decays in the scenario
in which PA dominates these amplitudes. On the other hand,
this relation is unknown in rescattering, which involves
long-distance contributions. For this reason, in this paper
we consider only PA.

We note that the transverse ($A_{\| ,\perp }$) and helicity
amplitudes ($A_\pm$) are related by $A_{\| ,\perp }=\left(A_+
\pm A_- \right)/\sqrt{2}$. However, $A_-$ for $B$ decays
(${\bar A}_+$ for ${\bar B}$ decays) has an extra spin-flip
$O(1/m_b)$ suppression with respect to $A_+$ (${\bar
A}_-$). Consequently, we neglect $A_-$ (${\bar A}_+$) and
henceforth define $|A_{\sss T}|^2 \equiv |A_\||^2 +
|A_\perp|^2 \approx |A_+|^2$ (i.e.\ $A_{\sss T}$ includes
both transverse amplitudes).

First, consider the pair of decays $\bs \to \phi \phi$ and
$\bd \to \phi K^{0*}$. The main PA contributions to the
transverse amplitudes are (the penguin-annihilation term
arises only from penguin operators with an internal $\bar t$
quark)
\bea
{\cal A}_{\sss T}(\bs \to \phi \phi) & = &
\left\vert V_{tb}^* V_{ts} \right\vert
f_{\sss \bsbar}f_{\sss \phi}^2 
~~ 2 (b_3^{\sss (\phi \phi)}+b_4^{\sss (\phi \phi)}), \nonumber \\
{\cal A}_{\sss T}(\bd \to \phi K^{0*}) & = &
\left\vert V_{tb}^* V_{ts} \right\vert
f_{\sss \bdbar}f_{\sss \phi}f_{\sss K^{0*}}
~~ b_3^{\sss (\phi K^{0*})},
\label{decays}
\eea
where $b_3^{(V_1V_2)}$ and $b_4^{(V_1V_2)}$ are the QCDf
terms corresponding to PA \cite{Benekeetal}. Throughout the
paper, we have dropped the overall factor of $G_{\sss
F}/\sqrt{2}$. (Absolute values are taken for the
Cabibbo-Kobayashi-Maskawa (CKM) matrix elements because we
are not interested in CP-violating observables here, but
rather the rate.)

As noted earlier, QCDf also contains the (perturbative
rescattering) term $\alpha_4$ which can contribute here. It
is the coefficient of the $(V-A) \otimes (V-A)$ piece in the
operator product expansion. However, the effect of $\alpha_4$
could be small in the transverse amplitude because of spin
flips. Consequently, it is reasonable to ignore $\alpha_4$ in
pure penguin decays, and we do so here.

The annihilation coefficient $b_3$ is given by
\cite{Benekeetal}
\bea
b_3^{\sss (V_1V_2)}=\frac{C_{\sss F}}{N_c^2}\bigl[
(C_5+ N_cC_6)\;A_3^f + C_5\;A_3^i +C_3\;A_1^i
\bigr] ~,
\label{b3}
\eea
where $N_c$ is the number of colors, the $C_i$ are Wilson
coefficients, and the incalculable infrared divergences are
found in the $A_{k}^{i,f}$ (due to the endpoint singularity
of the final-state distribution amplitudes). The superscripts
`$i$' and `$f$' refer to gluon emission from the initial- and
final-state quarks, respectively. The subscript `1' refers to
the Dirac structure $(V-A)\otimes(V-A)$, while `3' refers to
the Dirac structure $(-2)(S-P)\otimes(S+P)$.

The annihilation coefficient $b_4$ is 
\beq
b_4^{\sss (V_1V_2)}=\frac{C_{\sss F}}{N_c^2} \left(
C_4\;A_1^i + C_6\;A_2^i\right).
\label{b4}
\eeq
Here, $A_2^i\approx A_1^i$ has a suppression factor of
$O(1/m_b^2)$ compared to the $A_3^{i,j}$ \cite{Benekeetal}.
Thus, $b_4^{\sss (V_1V_2)}$ and the third term in $b_3^{\sss
(V_1V_2)}$ can be neglected in Eq.~(\ref{decays}).

Now, if we assume that the term containing $A_3^f$ in
$b_3^{\sss (V_1V_2)}$ dominates over the others (as we will
see below, the error on this approximation is at the level of
only a few percent), we find that
\beq
\frac{{\cal A}_{\sss T}(\bs \to \phi \phi)}
{2 ~ {\cal A}_{\sss T}(\bd \to \phi K^{0*})}=
\frac{f_{\sss \bs}}{f_{\sss \bd}} 
\frac{f_{\sss \phi}}{f_{\sss K^{0*}}} 
\frac{A_3^{\sss f(\phi \phi)}}{A_3^{\sss f(\phi K^{0*})}}  ~,
\label{ratio}
\eeq
where $A_3^{\sss f(V_1 V_2)}$ has the following integral form
\cite{Benekeetal}:
\bea
\label{int}
A_{3}^{\sss f(V_1 V_2)}&=&\pi\alpha_s \times \\ 
& & \hskip-2truecm \int^{1}_{0}dxdy \Bigl\{
\frac{2m_1}{m_2}r_\chi^{\sss V_2}
\frac{\phi_{a1}(x)\phi^{\perp}_{2}(y)}{x \bar y^2} +
\frac{2m_2}{m_1}r_\chi^{\sss V_1}\frac{\phi^\perp_{1}(x)\phi_{b2}(y)}
{x^2 \bar y}
\Bigr\}. \nn
\eea
Here $\phi^{\perp}$ ($\phi_{a,b}$) is the twist-2 (twist-3)
light-cone distribution amplitude, $x$ ($y$) stands for the
momentum fraction carried by the quark in $V_1$ ($V_2$), and
$r_\chi^{\sss V}=(2m_{\sss V}/m_b)f_{\sss V}^\perp/f_{\sss
V}$. $f_{\sss V}^\perp$ is defined as \cite{BeNe}
\beq
\bra{V(p,\epsilon^*)} {\bar q} \sigma_{\mu \nu} q' \ket{0} =
f_{\sss V}^\perp (p_\mu \epsilon^*_\nu - p_\nu
\epsilon^*_\mu) ~.
\eeq

It is useful to make a comment concerning the appropriate
interpretation of Eq.~(\ref{int}): if a ${\bar K}^{*}$ is one
of the final particles, the argument of the distribution
amplitudes (DAs) is the momentum fraction corresponding to
the $s$ quark, whereas for a $K^*$ meson, it is the momentum
fraction of the $d$ quark.  That is, the DAs are defined as
$\phi_{\sss K^*}(z)=\phi_{\sss {\bar K}^{*}}(\bar z)$, where
$\bar z \equiv 1-z$.

In order to estimate the ratio of the transverse amplitudes
in Eq.~(\ref{ratio}), we need to know the amount of SU(3)
breaking in the ratio of $A_3^{\sss f(\phi \phi)}$ and
$A_3^{\sss f(\phi K^{0*})}$.  To do this, further assumptions
are necessary.  In what follows, we adopt the same
assumptions as those used by the authors of
Ref.~\cite{Benekeetal} to carry out their study:

\begin{enumerate}

\item the asymptotic form of the light-cone distribution
(LCD) amplitudes,

\item a universal parametrization of the end-point
singularities, i.e.\ independent of any particular decay
mode,

\item a modeling of the singularities.

\end{enumerate}
It is the first point about which there has been some debate.
Certain references have calculated higher-order moments in
the LCDs, suggesting that such ``non-asymptotic LCDs'' are
important for some light mesons \cite{nonasymLCDs}. If so,
then SU(3) breaking in these non-asymptotic pieces will also
contribute to the ratio in Eq.~(\ref{ratio}).  Unfortunately,
this SU(3) breaking is not calculable, in which case our
analysis below will not hold.  Equally unfortunately, it is
very difficult for experiment to determine which type of LCD
is present \cite{exptLCD}. Thus, the reader should be aware
that our predictions are not only a test of PA dominance, but
also of asymptotic LCDs.

Now, the DAs for the final states are universal in the
asymptotic limit, i.e. $\phi^\perp(x)=6x\bar x$ and
$\phi_{a}(x)=\phi_{b}(\bar x)=3\bar x^2$. Thus, in this
approximation, we find that $A_3^{\sss f(\phi \phi)}$ and
$A_3^{\sss f(\phi K^{0*})}$ have exactly the same dependence
on $x$ and $y$ [Eq.~(\ref{int})]. In this case, the ratio
between the transverse amplitudes becomes
\bea 
\hskip-0.5truecm \frac{{\cal A}_{\sss T}(\bs \to \phi \phi)} 
{2 ~ {\cal A}_{\sss T}(\bd \to \phi K^{0*})} &=&
\frac{f_{\sss \bsbar}}{f_{\sss \bdbar}} 
\frac{f_{\sss \phi}}{f_{\sss K^{0*}}} \; \frac{2r_\chi^\phi}{\bigl(
\frac{m_{\sss K^{0*}}r_\chi^\phi}{m_\phi}+\frac{m_\phi
r_\chi^{\sss K^{0*}}}{m_{\sss K^{0*}}} \bigr)},
\label{ratio2}
\eea  
where $f_{\sss \bsbar}/f_{\sss \bdbar} = 1.22 \pm 0.03$
\cite{lat}, $f_{\phi} = 221 \pm 3~{\rm MeV}$ \cite{BeNe}, and
$f_{\sss K^{0*}} = 218 \pm 4~{\rm MeV}$ \cite{BeNe}. The
values of $f_{\sss V}^\perp$ in $r_\chi^{\sss V}$ are
theoretically estimated.  We have \cite{BeNe} $f_{\sss
\phi}^\perp = f_{\sss K^*}^\perp = 175 \pm 25$ {\rm MeV}.
Since the decay constants and the meson masses are known, the
SU(3) breaking in Eq.~(\ref{ratio2}) is well-controlled for
this pair.

What this says is that the transverse polarization amplitude
in $\bs \to \phi \phi$ is predicted by PA to be related to
that in $\bd \to \phi K^{0*}$ through Eq.~(\ref{ratio2}).
Thus, once one makes the polarization measurement in the
$\bd$ decay, one can test PA by making the equivalent
measurement in the $\bs$ decay.

The key ingredient in the above analysis is to take two
decays in which the final states have the same dependence on
the momentum fractions $x$ and $y$.  However, although the
pair of decays considered above is the most promising for the
analysis, it is not unique.  In fact, all decays in a special
class have the same dependence on $x$ and $y$.  This class
contains $\bar b (\bd, B^{\sss +}) \to \bar s d \bar d $ (the
parenthesis indicates that this transition includes only
$\bd$ or $B^{\sss +}$, and not $\bs$, decays) and $\btos s
\bar s$ penguin decays.  In addition, only decays to
ground-state spin-1 mesons are included (excited mesons have
different DAs in general). Thus, excluding those decays which
also receive tree contributions, these correspond to $B^{\sss
+} \to \rho^{\sss +} K^{0*}$, $\bd \to \phi K^{0*}$, $B^{\sss
+} \to \phi K^{{\sss +}*}$, and $\bs \to \phi \phi$.  The
important point here is that the dependences on the momentum
fractions in $A_3^f$ are the same for {\bf every} decay
belonging to this class. Therefore, in the comparison of any
two of these decays, the integrals containing the
singularities cancel in the ratio of the transverse
amplitudes, and this even before using a cutoff to regulate
the end-point divergences.

Of course, the size of the SU(3) breaking will depend on the
pairs of decays considered [see Eq.~(\ref{ratio2})]. For
example, we expect the pair $\bs \to \phi \phi$ and $B^{\sss
+} \to \phi K^{{\sss +}*}$ to be as good as $\bs \to \phi
\phi$ and $\bd \to \phi K^{0*}$.  However, the SU(3) breaking
in $B^{\sss +} \to \rho^{\sss +} K^{0*}$ and $B^{\sss +} \to
\phi K^{{\sss +}*}$ also turns out to be about the same
size. In all cases, the SU(3) breaking can be worked out as
we have done above.

Another class in which the final states have the same
dependence on $x$ and $y$ is given by the decays $\bar b
(\bs) \to \bar d d \bar d$ and $\bar b (\bs) \to \bar s d
\bar d$. Even if one of the most promising $\bs$ decay modes
to be measured in the near future, $\bs \to K^{0*} {\bar
K}^{0*}$, belongs to this class, we cannot use it to make
predictions because its decay-class partner, $\bs \to \rho^0
{\bar K}^{*0}$, typically receives tree
contributions. Therefore, this class is not very useful since
it only contains one pure penguin decay.

The third class is defined by the transition $\btod s \bar
s$.  This includes the decays $\bs \to \phi {\bar K}^{0*}$
and $\bd \to {\bar K}^{0*} K^{0*}$, and this pair is
particularly promising. The prediction for the ratio of the
transverse amplitudes is
\bea 
\hskip-0.2truecm
\frac{{\cal A}_{\sss T}(\bd \to {\bar K}^{0*} K^{0*})} {{\cal A}_{\sss T}
(\bs \to \phi {\bar K}^{0*})} =
\frac{f_{\sss \bdbar}f_{\sss K^{0*}}}{f_{\sss \bsbar}
f_{\sss\phi}}\; \frac{2r_\chi^{\sss K^{0*}}}{\bigl(
\frac{m_{\sss K^{0*}}r_\chi^\phi}{m_\phi}+\frac{m_\phi
r_\chi^{\sss K^{0*}}}{m_{\sss K^{0*}}} \bigr)} .
\label{ratio3}
\eea 
The pair $\bs \to \phi {\bar K}^{0*}$ and $B^{\sss +} \to
K^{{\sss +}*} K^{0*}$ can be treated similarly.

An important consequence of the above discussion is that it
is not possible to get cancellations in the $A_3^f$ ratios
through the comparison of decays belonging to different
classes. For example, we will not obtain a clean result by
comparing $\bs \to K^{0*} {\bar K}^{0*}$ and $\bd \to {\bar
K}^{0*} K^{0*}$, even though the final states are identical
in the two decays. In addition, we have presented the list of
decays in Eq.~(\ref{list}) in terms of $\btod$ or $\btos$
transitions, motivated by the apparent predominant role
played by the $\btos$ transitions in the polarization puzzle.
However, as shown above, this is not the most natural way to
classify the decays in order to achieve the most reliable
predictions for penguin annihilation within QCDf.

We now turn to the estimation of errors in
Eqs.~(\ref{ratio2}) and (\ref{ratio3}) due to the single
inclusion of $A_3^f$ in the transverse amplitudes. We study
the relative magnitude of the $A_{1,3}^i$ terms in $b_3$
[Eq.~(\ref{b3})], as well as the relevance of the
annihilation coefficient $b_4$ [Eq.~(\ref{b4})] in the cases
it is appropriate.  To do this, we follow several
applications of QCDf in which the incalculable infrared
divergences in the $A_{k}^{i,f}$ are isolated with a cutoff
and the penguin annihilation amplitude is modeled by
introducing unknown parameters.

In passing, we note the following.  Previously, we mentioned
that we use the same assumptions as those in
Ref.~\cite{Benekeetal}.  Although we work within the same
restricted theoretical framework as this reference, and
although it is true that the size of the error in our
predictions could be affected by large uncertainties related
to the choice of this particular scenario, we emphasize that
our results go beyond the analysis made in
Ref.~\cite{Benekeetal}. There, due to the parametrization of
the infinities, the uncertainties in the individual
transverse amplitudes turn out to be at the level of one
hundred percent for most of the decay channels
\footnote{The authors in Ref.~\cite{Benekeetal} overcome this
difficulty by taking into account experimental data to fit
the entire divergent transverse amplitude (i.e., ${\hat
\alpha}^{c {\sss -}}_4$ in their notation).}. Instead, we
show here that it is possible to obtain more accurate
predictions with the same theoretical inputs used in the
treatment of the divergent integrals when specific decays are
compared.

To illustrate the procedure used in the estimation of errors,
we focus on the first pair, $\bs \to \phi \phi$ and $\bd \to
\phi K^{0*}$ (a similar reasoning holds for $\bs \to \phi
{\bar K}^{0*}$ and $\bd \to {\bar K}^{0*} K^{0*}$, as well as
several other decay pairs). We first write the
$A_{1,3}^{f,i}$ after the parametrization of the infrared
divergences is applied \cite{Benekeetal}
\bea & &\hskip-0.95truecm A_{1,2}^i \approx 18\pi\alpha_s\;
\frac{m_1m_2}{m_{\sss B}^2} \bigl(\frac{1}{2}X_{\sss L}+\frac{5}{2}
-\frac{\pi^2}{3}\bigr), \nonumber \\ & &\hskip-0.7truecm A_3^i \approx
18\pi\alpha_s\; \bigl(\frac{m_1}{m_2}r_\chi^{\sss
V_2}-\frac{m_2}{m_1}r_\chi^{\sss V_1}\bigr)\; (X_{\sss A}^2-2X_{\sss
A}+2), \nonumber \\ & &\hskip-0.7truecm A_3^f \approx 18\pi\alpha_s\;
\bigl(\frac{m_1}{m_2}r_\chi^{\sss V_2}+\frac{m_2}{m_1}r_\chi^{\sss
V_1}\bigr)\; (2X_{\sss A}^2-5X_{\sss A}+3), \eea
We see that $A_{1,2}^i$ and $A_3^f$ are symmetric in the
interchange of $V_1 \leftrightarrow V_2$, while $A_3^i$ is
antisymmetric. $X_{\sss A}$ and $X_{\sss L}$ contain the same
input parameters, but have different end-point-divergence
behavior:
\bea
& & X_{\sss A}= (1+\rho_{\sss A}e^{i\phi_{\sss
    A}})\ln\frac{m_{\sss B}}{\Lambda_h} ~,
\nonumber \\
& & X_{\sss L}=(1+\rho_{\sss A}e^{i\phi_{\sss
A}})\frac{m_{\sss B}}{\Lambda_h} ~.
\eea
Here, $\Lambda_h$ is an input parameter ($\Lambda_h=0.5$ GeV
\cite{Benekeetal}), and $\phi_{\sss A}$ is an arbitrary
phase. We have taken $\Lambda_h$, $\phi_{\sss A}$, and
$\rho_{\sss A}$ to be the same for every decay mode.

To study the relative significance of $b_4$ and the neglected
terms in $b_3$, we evaluate the following ratios
[Eqs.~(\ref{b3}) and (\ref{b4})]:
\bea
& & r_{b_4}^{\sss (V_1 V_2)} = \frac{C_4 + C_6}{C_5 + N_c C_6}\; 
\frac{A_{1,2}^{\sss i(V_1 V_2)}}{A_3^{\sss f(V_1 V_2)}} \;, \nn \\
& & r_{b_3}^{\sss (V_1 V_2)} = \frac{C_3}{C_5 + N_c C_6}\; 
\frac{A_1^{\sss i(V_1 V_2)}}{A_3^{\sss f(V_1 V_2)}} \;, \nn \\
& &\hskip-0.15truecm R_{b_3}^{\sss (\phi K^{0*})} = 
\frac{C_5}{C_5 + N_c C_6}\; 
\frac{A_3^{\sss i(\phi {\bar K}^{0*})}}{A_3^{\sss f(\phi {\bar K}^{0*})}} \;,
\label{3r}
\eea
where $V_1 V_2 = \phi \phi$, $\phi K^{0*}$ (note that
$R_{b_3}^{(\phi \phi)}$ is zero).

Although the values of these ratios are quite uncertain, in
large part due to the (arbitrary) value of $\Lambda_h$, in
virtually all cases it is found that $|r_{b_4}^{\sss (V_1
V_2)}| \lesssim {\cal O}(\mbox{10}^{-2})$ and $|r_{b_3}^{(V_1
V_2)}|, |R_{b_3}^{(\phi K^{0*})}| \lesssim {\cal
O}(\mbox{10}^{-3})$. One can see this as follows. First, the
ratios of the relevant Wilson coefficients in Eq.~(\ref{3r})
at $\mu=m_b/2$ are as follows \cite{BBNS}:
\bea
& & \hskip2truecm \frac{C_4 + C_6}{C_5 + N_c C_6}\approx 0.63 ~, \nn \\
& & \frac{C_3}{C_5+N_c C_6}\approx -0.11 ~,~~
\frac{C_5}{C_5+N_c C_6}\approx -0.05 ~, 
\eea
Second, we have $(A_1^{\sss i(V_1 V_2)}/A_3^{\sss f(V_1
V_2)})\sim O(m_1m_2/m_{\sss B}^2)$, and $|A_3^{\sss i(\phi
K^{0*})}/A_3^{\sss f(\phi K^{0*})}|\sim |(a-b)/(a+b)|\approx
0.07$ (where $a$ denotes $(m_{\sss
K^{0*}}/m_\phi)r_\chi^{\sss \phi}$ and $b$ is given by
$K^{0*} \leftrightarrow \phi$).

We evaluate the three ratios in Eq.~(\ref{3r}) by considering
many different values in the ranges $0 \leq \rho_{\sss A}
\leq 2$ and $0 \leq \phi_{\sss A} \leq 2\pi$. We find that
$|r_{b_4}^{\sss (V_1 V_2)}|$, $|r_{b_3}^{\sss (V_1 V_2)}|$,
$|R_{b_3}^{\sss (\phi K^{0*})}| \ll 1$ always, except for a
singular behavior at $\phi_{\sss A}=0, 2\pi$. The largest
contribution to the error arises from $r_{b_4}^{\sss (V_1
V_2)}$ but it remains at the level of a few percent within
the scanned region of the parameter space. Thus, we have
covered a wide set of models of the infrared singularities.
The point here is that, although the precise values of the
ratios are very uncertain, they are always small.

We therefore conclude that the PA dominance hypothesis leads
to a clean prediction for the ratio of transverse amplitudes
in the pair $\bs \to \phi \phi$ and $\bd \to \phi K^{0*}$
[Eq.~(\ref{ratio2})], though this result is a direct
consequence of the particular modeling of the suppressed
terms (e.g.\ asymptotic LCDs). We have also analyzed the
pair $\bs \to \phi {\bar K}^{0*}$ and $\bd \to {\bar K}^{0*}
K^{0*}$, as well as the pairs containing the charged modes
and ($B^{\sss +} \to \rho^{\sss +} K^{0*}$, $B^{\sss +} \to
\phi K^{{\sss +}*}$), under the same set of assumptions, and
we have obtained similar conclusions.

We have therefore seen that there are a number of decay pairs
within a given class whose transverse polarizations are
related. Some of these pairs involve a $\bs$ decay. Now, the
$B$-factories BaBar and Belle have made many measurements of
$\bd$ and $B^{\sss +}$ mesons. But it is only relatively
recently, at hadron colliders, that $\bs$ mesons have started
to be studied. This will increase when the LHCb turns on. It
should be possible to make measurements of the transverse
polarization in some $\bs$ decays in the near future, and to
test the PA/QCDf hypothesis.

Above, we have presented the ratio of ${\cal A}_{\sss T}$'s
for two pairs of decays.  However, it is perhaps better to
present a ratio of $\fT$'s since this is what will actually
be measured. ${\cal A}_{\sss T}$ and $\fT$ are related by
including information about the branching ratio ($BR$): $\fT
= |{\cal A}_{\sss T}|^2/(\Gamma~ BR/PS)$, where $\Gamma$ is
the total width and $PS$ is the phase space.  We find
\bea
{\fT(\bs \to \phi \phi) \over \fT(\bd \to \phi
K^{0*})} & = & \nn\\
& & \hskip-1.4truecm 3.22 \pm 0.72 \, {BR(\bd \to \phi K^{0*}) 
\over BR(\bs \to \phi \phi)} ~, \nn\\
{\fT(\bd \to {\bar K}^{0*} K^{0*}) \over \fT(\bs \to \phi {\bar K}^{0*})} & = &
\nn\\
& & \hskip-1.4truecm 0.62 \pm 0.12 \, {BR(\bs \to \phi {\bar K}^{0*}) \over 
BR(\bd \to
{\bar K}^{0*} K^{0*})} ~.
\label{predictions}
\eea
The numbers have been obtained by taking values of masses and
lifetimes from the Particle Data Group without errors
\cite{pdg}, along with the theoretical estimates of the decay
constants given above. The predictions given in
Eq.~(\ref{predictions}) will yield a test of PA. If there are
discrepancies in the measurements, this may indicate the
presence of new physics, in $\btos$ and/or $\btod$
transitions.

The decays $\bd \to {\bar K}^{0*} K^{0*}$ \cite{K0*K0bar*}
and $\bd \to \phi K^{0*}$ \cite{phiK0*recent} have both been
measured, so that this information can be included in
Eq.~(\ref{predictions}):
\bea
\bd \to {\bar K}^{0*} K^{0*} & : & \fL =
0.81^{+0.10}_{-0.12} \pm 0.06 ~, \nn\\
& & BR = (0.96^{+0.21}_{-0.19}) \times 10^{-6} ~. \nn\\
\bd \to \phi K^{0*} & : & \fL =
{ 0.49}\pm{0.04} ~, \nn\\
& & BR = (9.5 \pm 0.8) \times 10^{-6} ~.
\eea
$\fT$ is defined as $\fT = 1 - \fL$. 

To summarize, a large $\fTfL$ has been observed in several
$\btos$ $B \to V_1V_2$ decays. There are two
explanations of this measurement within the standard model --
penguin annihilation (PA) and rescattering. Now, one
logically also expects to see a large $\fTfL$ in certain
$\btod$ decays.  The most promising decays are those
dominated by penguin amplitudes, and there are quite a few
$\btod$ and $\btos$ penguin decays.  All of these are equal
under flavor SU(3) symmetry. Given the measurement of $\fTfL$
in one decay, if one wishes to predict $\fTfL$ in another
decay, it is necessary to take SU(3) breaking into
account. However, it is only within a specific scenario of PA
dominance for the transverse amplitudes, and for special
classes of decay pairs, that this SU(3) breaking can be
estimated. We therefore assume that it is PA alone which is
the source of the large transverse polarization and explore
its consequences.

We find that there are several decay pairs for which PA makes
a reasonably precise estimate of the SU(3) breaking (assuming
asymptotic LCDs).  Thus, given the measurement of $\fTfL$
in one decay, PA makes a prediction for the transverse
polarization in the second decay. In this paper we have
concentrated on two decay pairs that involve $\bs$ mesons:
($\bs \to \phi \phi, \bd \to \phi K^{0*}$) and ($\bs \to \phi
{\bar K}^{0*}, \bd \to {\bar K}^{0*} K^{0*}$). The
polarization measurement in the $\bd$ decay allows one to
predict the transverse polarization in the $\bs$ decay.  This
will permit the explicit testing of PA, probably in the near
future at the LHCb.

\bigskip
\noindent
{\bf Acknowledgments}:
This work was financially supported by NSERC of Canada (DL,
MN \& AS), and by FPA2005-02211, PNL2005-51 and the Ramon y
Cajal Program (JM).


\bibliographystyle{h-physrev2-original}

\end{document}